# Anomalous Magnetic Behavior in $Ba_2CoO_4$ with Isolated $CoO_4$ Tetrahedra

Qiang Zhang[1,2], Guixin Cao[1], Feng Ye[2], Huibo Cao[2], Masaaki Matsuda[2], D. A. Tennant[2], Songxue Chi[2], S. E. Nagler[2], W.A. Shelton[3], Rongying Jin[1], E. W. Plummer[1], and Jiandi Zhang[1*]

[1]Department of Physics and Astronomy, Louisiana State University, Baton Rouge, Louisiana 70803, USA

[2]Neutron Scattering Division, Oak Ridge National Laboratory, Oak Ridge, Tennessee 37831, USA

[3]Cain Department of Chemical Engineering, Louisiana State University, Baton Rouge, Louisiana 70803, USA

## Abstract

The dimensionality of the electronic and magnetic structure of a given material is generally predetermined by its crystal structure. Here, using elastic and inelastic neutron scattering combined with magnetization measurements, we find unusual magnetic behavior in three-dimensional (3D) $Ba_2CoO_4$. In spite of isolated $CoO_4$ tetrahedra, the system exhibits a 3D noncollinear antiferromagnetic order in the ground state with an anomalously large Curie-Weiss temperature of 110 K compared to $T_N$ = 26 K. More unexpectedly, spin dynamics displays quasi-2D spin wave dispersion with an unusually large spin gap, and 1D magnetoelastic coupling. Our results indicate that $Ba_2CoO_4$ is a unique system for exploring the interplay between isolated polyhedra, low-dimensional magnetism, and novel spin states in oxides.

It is anticipated that low-dimensional magnetism in a material is directly related to its low dimensional crystal structure [1], such as in cuprates [2], Fe-based pnictides [3], ferromagnetic semiconductor $CrSiTe_3$ [4], double-layer perovskite $Sr_3(Ru_{1-x}Mn_x)_2O_7$ [5], or Weyl semimetals in $YMnBi_2$ [6], etc. In these materials, the quasi-two-dimensional (2D) magnetism originates from the clearly layered crystal structure where the nearest-neighbor (NN) M-M distance (where M is a transition metal element) along the interlayer direction is much larger than that within the layer, which yields a very weak interlayer magnetic interaction [1-6]. It would be of great interest to explore whether quasi-2D magnetism could be realized in a nonlayered compound involving comparable NN M-M distances in a crystallographically 3D system. The identification of such a system may shed light on the microscopic origin of quansi-2D magnetism.

Different from other well-studied cobaltates with the Co ion in an octahedral environment, monoclinic $Ba_2CoO_4$ [7-8] has a 3D crystallographic structure with *isolated* tetrahedral $CoO_4$ without any corner-, edge- or face-sharing [Fig. 1(a)]. Naively one would expect that the spin correlation in $Ba_2CoO_4$ is very weak, presumably dictated by spin dipole-dipole interaction with energy $U \approx \frac{1}{(137)^2}\left(\frac{a_0}{4}\right)^3 Ry$ [9]. Using the shortest Co-Co distance of 4.67 Å, we estimate $U \sim$ 0.01 meV, implying that any magnetic ordering would happen below 0.1 K. Yet, $Ba_2CoO_4$ exhibits AFM ground state below $T_N \approx 26$ K [8,10,11] with an anomalously high Curie-Weiss temperature ($|\theta| \sim 110$ K), which is > 4 times larger than $T_N$ [11]. The large $|\theta|/T_N$ ratio could be a result of spin frustration [12, 13] or low dimensional magnetism [14] existing in the system. To complicate any interpretation of magnetism is the fact that the reported magnetic structures are inconsistent with each [10, 15]. Boulahya *et al.* [10] reported a canted AFM order in the *bc* plane based on powder neutron diffraction measurements. In contrast, muon spin rotation and relaxation ($\mu^+$SR) experiments [15] found that the magnetic moment is basically along the *a* axis. In addition, spin dimer analysis [16] for the magnetic coupling in $Ba_2CoO_4$ has no indication of quasi-2D magnetism. Super-superexchange (SSE) mechanism [16-19], which describes spin interactions beyond direct Co-O-Co superexchange pathways, was proposed to be responsible for the magnetic interaction in $Ba_2CoO_4$. However, there is no experimental confirmation to date.

Here, we demonstrate different dimensionality between static and dynamical magnetism in $Ba_2CoO_4$. The system exhibits a 3D noncollinear AFM order with buckled zig-zag chains along the *b* axis below $T_N$ = 26 K but 1D magnetoelastic coupling occurs along the *a* direction, despite well-separated $CoO_4$ tetrahedra [Fig. 1(a)]. However, the spin waves (SWs) display a quasi-2D character with dispersion in the *ab*-plane. An anomalous spin gap (~2.55 meV) comparable to the SW bandwidth reflects a large magnetic anisotropy. The magnon dispersion relation analyzed using the linear SW theory reveals large anisotropic magnetic interactions. The results can be interpreted in terms of a frustrated network of Co-O···O-Co spin exchange pathways where the overlapping oxygen *p*-orbitals determine the amplitude of magnetic interactions. The uniaxial magnetoelastic effect is the evidence of a certain spin-lattice coupling to stabilize the 3D AFM order against the spin frustration.

$Ba_2CoO_4$ crystals were synthesized using the floating zone method [20]. Single-crystal neutron diffraction was used to determine the structure of $Ba_2CoO_4$ [21,22], revealing a monoclinic structure with space group *$P2_1/n$* (No. 14) at 5 K as illustrated in Fig. 1(a). There are 4 Co atoms in one crystalline unit cell. This result is consistent with previous neutron and x-ray diffraction measurements [10,11]. Temperature (*T*)-dependence of the magnetization (M) for $Ba_2CoO_4$ in a field of 0.1 T is shown in Fig. 1(b) for three principal directions. Note that M decreases with increasing temperature above $T_N$ with no anisotropy. Below $T_N$ = 26 K, $M_a$ drops much faster than $M_c$, while $M_b$ increases after a small drop. This indicates anisotropic magnetism below $T_N$, with the AFM configuration along both the *a* and *c* directions, but ferromagnetic (FM)-like alignment along the *b* direction. The rapid decrease of $M_a$ below $T_N$ implies that the moment direction mainly points to the *a* axis. Fitting to the inverse susceptibility ($H//a$) at the high-temperature linear portion of the curve with the Curie-Weiss law ($\chi = \frac{C}{T+\theta}$) yields a Curie-Weiss temperature $\theta \approx 109$ K, consistent with the previous reports [10,11].

Figures 1(c) and (d) show the temperature dependences of the lattice constants determined by measuring the *Q* scans through the nuclear peaks (400), (020) and (004) of neutron diffraction, given there is a negligible change in the monoclinic beta angle <0.6° between RT and 5 K. Above $T_N$, the lattice constants in all three directions show identical *T* dependence with thermal expansion coefficient $\alpha \sim 9 \times 10^{-6}$ K$^{-1}$ (close to the value for glass of $7.6 \times 10^{-6}$ K$^{-1}$ [23]). However, the lattice constant *a* exhibits anomalous behavior with an abrupt and nonlinear drop below $T_N$, indicating strong magnetoelastic coupling in this specific direction. The *T*-dependence of the lattice constant *a* (Fig. 1(c)) scales inversely with the magnetic order parameter shown in Fig. 2(a).

Magnetic peaks with a commensurate propagation wavevector $k$ = (-1/2, 0, 1/2) appear below $T_N$ in the neutron diffraction due to long range magnetic ordering. The *T*-dependent peak intensity of the magnetic Bragg peak (-1/2, 1, 1/2) in Fig. 2(a) shows an AFM transition at $T_N$. As illustrated in Fig. 2(b), the overall magnetic structure is AFM with an ordered moment of 2.69(4) $\mu_B$/Co, consisting of 16 Co spins in the magnetic unit cell. The magnetic unit cell is $2a \times b \times 2c$ with respect to the crystalline unit cell with $|m_a|$ = 2.377 $\mu_B$, $|m_b|$ = 1.128 $\mu_B$, and $|m_c|$ = 0.586 $\mu_B$, respectively. All the Co spins are antiparallel along the *a* and *c* axes, but double-stripe type parallel configuration along the *b*-axis, consistent with the magnetization measurements. The moments primarily point along the *a* axis and alternatively canted with a canting angle to the *b* axis ~ ±25° and to the *c* axis ~ ±13° [21,24,25].

The spin dynamics is investigated using inelastic neutron scattering. The inset of Fig. 2(a) shows the spectra of the constant-*Q* energy scan at two magnetic zone centers, (1/2, 0, 1/2) and (1/2, 1, 1/2) at 5 K. A spin gap of Δ ≈ 2.55(3) meV is observed at 5 K and disappears at $T_N$, confirming its magnetic origin. This is in stark contrast to the isotropic nature of the high spin state $Co^{4+}$ (*S*=5/2, *L*=0, $e^2t_2^3$) reported previously [7-8,10-11,15] due to the full quenching of the orbital angular momentum and consequently the absence of any spin gap. The AFM ordered moment of 2.69(4) $\mu_B$ is also much lower than 5$\mu_B$ expected for high spin state of $Co^{4+}$ but close to 3$\mu_B$ for the intermediate spin state of $Co^{4+}$ (*S*=3/2). Both indicate that the ground state of $Ba_2CoO_4$ is in the intermediate spin (IS) state $Co^{4+}$ (*S*=3/2, *L*≠0, $e^3t_2^2$). While the IS state of $Co^{4+}$ was frequently observed in cobaltites when Co is in an octahedral environment, the IS state of $Co^{4+}$ in the tetrahedral environment seems very rare. Early theoretical calculations [26] indicated that all the IS (*S*=3/2) states of $3d^5$ cation in the tetrahedral environment may not be stable. Recently, Kauffmann et al. [27] proposed that the off-centering of O atoms from its ideal tetrahedral positions may induce the intermediate spin state of tetrahedral $Co^{4+}$ ion. The x-ray absorption spectroscopy and/or theoretical calculations are needed, in order to investigate the microscopic origin of the possible IS state in $Ba_2CoO_4$.

Figures 2(c-f) display the contour plots of *S*(*Q*, *E*), determined from the constant-*Q* energy scans along the [*H* 0 0], [0 *K* 0], [0 0 *L*] and diagonal [*H* 0 *H*] direction at 5 K (see Note 3 in SM). Two SW branches appear along the [*H* 0 0] and [*H* 0 *H*] directions with the more dispersive being the high-energy branch. These two branches nearly merge along the [0 *K* 0] and [0 0 *L*] directions. Surprisingly, the magnon bandwidth is less than 3 meV, close to the spin gap value of ~2.6 meV.

In contrast, both the energy and intensity of the SW along the [0 0 L] direction show negligible dispersion, indicating that Ba$_2$CoO$_4$ exhibits a surprising quasi-2D magnetism.

To identify the observed SW modes and quantitatively obtain the magnetic exchange parameters, we have performed the linear SW calculations using the SpinW package [29] with an effective Heisenberg-like Hamiltonian given by

$$H = \sum_{i \neq j} J_{ij} \vec{S}_i \cdot \vec{S}_j + \sum_i A_i \vec{S}_i^2 \qquad (1)$$

where $\vec{S}_i$ denotes the spin of magnetic Co ion at site $i$, $J_{ij}$ describes the magnetic exchange coupling constant between spin-pairs at site $i$ and $j$, $\sum_{i \neq j}$ indicates summation over pairs of spins, and $A_i$ is the diagonal element of the 3×3 single-ion anisotropy matrix. The best fitting values of the exchange parameters $SJ_{ij}$ and anisotropy parameter $SA$ as listed in Table I. Compared to $SA_b$ and $SA_c$, the anisotropy parameter $SA_a$ is extremely small, consistent with the ordered moment mainly aligned to $a$ axis. Due to the large spin gap, it is impossible to fit the experimental data of spin waves without large anisotropy term. It is worthwhile noting that the spin dimer analysis proposed in Ref. 16 does not include single-ion anisotropy.

Figure 3(a) illustrates the magnetic structure, doubling of the $ab$ plenary lattice structure (PLS) and the important exchange constants used in the fitting of spin waves. Considering only Co sites for two stacking PLSs shown in Fig. 3(a), each Co atom is surrounded by six NN Co atoms with comparable Co-Co distances. The two adjacent PLSs are not identical, thus there are two quasi-2D PLSs in the magnetic unit cell alternating along the $c$ axis and connected by the exchange coupling of $J_\perp$. Figure 3(b) shows the projection of one quasi-2D PLS onto the $ab$ plane. The spin configuration within each PLS is collinear but the spin ordering between the adjacent PLSs is not, forming an overall noncollinear AFM order. For each quasi-2D PLS, the spins form a buckled

AFM double-chain structure with strong exchange couplings [$J_1$, and $J_1'$, see Fig. 3(b)] along the $b$ direction. Six NN pairs (including four intra-PLS pairs, $J_1$, $J_1'$, $J_2$, $J_2'$, and two equivalent inter-PLS pairs, $J_\perp$) and one next nearest neighbor (NNN) pair $J_3$, marked in Fig. 3(a-b), plus three diagonal elements ($A_i$) of single-ion anisotropy matrix, are included to fit the dispersion. Using the fitting $J$ values and the determined magnetic structure, the corresponding SW spectra $S(Q, E)$, along the four measured directions in the reciprocal space in units ($H,K,L$) are simulated as shown in Fig. 3(c-f), which are in good agreement with the experimental data in Fig. 2(c-f).

The important messages from the SW fitting and simulation are: 1) The inter-PLS $J_\perp$, which characterizes the dispersion along the $c$ axis is less than 5% of the intra-PLS $J_1$, $J_1'$, $J_2$, $J_2'$, and $J_3$, reflecting the quasi-2D magnetism. 2) Spins form AFM zigzag chains along the $b$ axis due to the strong magnetic interactions $J_1$ and $J_1'$. These chains are coupled through the unexpected strong NNN interaction $J_3$, comparable with $J_1$ and $J_1'$. 3) Within one PLS [see Fig. 3(b)], there are two distinct distorted triangles marked by grey ($J_1$, $J_2$ with $J_3$) and blue ($J_1'$, $J_2'$ with $J_3$), respectively. Although all the fitted $J_{ij}$ values are positive, i.e., AFM, the spin configurations associated with $J_2$ and $J_2'$ are FM [see Fig. 3(b)], therefore, this lifts spin frustration within these triangles and leads to long-range magnetic order. 4) The anisotropic coefficients in the $b$ and $c$ directions are three orders larger than that in the $a$ direction. This indicates that spins prefer to be aligned along the $a$ axis with a collinear configuration. Such energy more favorable state is obtained because of the 1D magnetoelastic effect [Fig. 1(c)]. The 1D magnetoelastic coupling along $a$ direction at $T_N$ may be associated with the details of the exchange interaction [30,31]. In this case, $J_3$ needs to be large enough to stabilize such a magnetic structure against spin frustration in two triangular lattices. The magnetoelastic effect below $T_N$ exactly reflects the correlation of $J_3$ and lattice constant in the $a$ direction.

The quasi-2D magnetism in $Ba_2CoO_4$ is novel and fundamentally different from the conventional quasi-2D magnetism compounds [2-6]: i) The crystal structure is 3D with isolated $CoO_4$ tetrahedra, lacking well separated magnetic and nonmagnetic layers as in conventional quasi-2D magnetism compounds; ii) There are two distinct stackings repeated along the *c* axis in $Ba_2CoO_4$. The spin arrangement is collinear within each of stacking, but they are noncollinear between these two stackings connected by a negligible $J_\perp$ to be responsible for the quasi-2D magnetism; iii) Although intra-PLS Co-Co distance of $J_\perp$ is shorter than those of the intra-PLS couplings $J_1'$, $J_2'$ and $J_3$, the magnitude of $J_\perp$ between the two stackings is two orders smaller.

The static and dynamic magnetic behavior of magnetism in $Ba_2CoO_4$ are schematically summarized in Fig. 4(a), 1D spin-lattice coupling, quasi-2D spin waves and 3D magnetic order. Given the fact that $CoO_4$ tetrahedra are isolated with large Co-Co spacing (> 4.662 Å), there should be little direct exchange interaction between Co atoms. Thus, the spin-spin interaction via indirect spin exchange pathways ought to be considered. For $Ba_2CoO_4$, indirect spin interactions may take place through Co-O···O-Co or Co-O-Ba-O-Co exchange paths. The exchange path of Co-O···O-Co is referred as the super-superexchange (SSE) mechanism [16,19], in contrast to conventional superexchange model [32]. The sign and the magnitude of interactions via Co-O···O-Co are not necessarily governed by the direct Co-Co distances, but by the overlap of orbitals along Co-O···O-Co, especially the overlap of their *p*-orbitals of the non-bonding O···O in the vicinity of the van der Waals distance [16-19]. Given the O···O distances and the Co-O···O or O···O-Co angles are critical in determining the overlap of the O *p*-orbitals, it is important to compare the obtained magnetic exchange parameters with the corresponding crystallographic parameters. Thus, we determined the O···O distances and the Co-O···O or O···O-Co angles based on our Rietveld refinements of neutron diffraction results. As shown in Figs. 4(b-d) and Ref. [33], the inter-PLS

Co-Co tetrahedra are less coplanar than intra-PLS Co-Co tetrahedra, which does not favor the overlap of the O $p$-orbitals. In the two exchange pathways Co-$O_1$···$O_3$-Co and Co-$O_1$···$O_2$-Co for $J_\perp$ [see Fig. 4(c)], the O···O distances are comparable to those for $J_1$', $J_2$' and $J_3$ [see Figs. 4(b) and (d)], but the ∠ Co-O···O and ∠O···O-Co angles are close to 90° and smaller than the corresponding angles for $J_1$', $J_2$' and $J_3$, thus leading to a much weaker $J_\perp$ based on SSE model (see Note 4 in SM for more details).

In summary, we have investigated magnetic structure, magnetic interactions and magnetoelastic coupling in $Ba_2CoO_4$ with isolated $CoO_4$ tetrahedra. The system exhibits a 3D noncollinear long-range AFM order below $T_N$ = 26 K with magnetic moment primarily along the $a$-axis. The spin excitation spectra reveal a quasi-2D SW dispersion with an unusually large spin gap ~ 2.55(3) meV and the $T$-dependent lattice constants clearly illustrate a strong 1D magnetoelastic effect along $a$-axis. The concurrence of 3D lattice structure, 3D noncollinear magnetic structure, quasi-2D spin waves dispersion, 1D magnetoelastic coupling, and the unusual intermediate spin state of $Co^{4+}$ in a tetrahedral environment make $Ba_2CoO_4$ a unique system for exploring novel magnetism. Our work may open a new avenue to investigate quasi-2D magnetism in nonlayered structure involving isolated coordinate polyhedron and could be an important stimulus to explore the very rare intermediate spin state in the $3d^5$ cations such as $Mn^{2+}$, $Fe^{3+}$, $Co^{4+}$ in the tetrahedral environment.

**Acknowledgments**: We would like to thank Zhentao Wang and Sándor Tóth for helpful discussions. This work was primarily supported by the U.S. Department of Energy under EPSCoR Grant No. DE-SC0012432 with additional support from the Louisiana Board of Regents. Use of the high flux isotope reactor at the Oak Ridge National Laboratory, was supported by the US Department of Energy, office of Basic Energy Sciences, Scientific User Facilities Division.

**Table I.** The optimal parameters obtained by fitting experimental SW dispersions with the linear spin wave theory (see details in the main text), compared with corresponding Co-Co distances.

|  | $SJ_1$ | $SJ_1'$ | $SJ_2$ | $SJ_2'$ | $SJ_\perp$ | $SJ_3$ | $SA_a$ | $SA_b$ | $SA_c$ |
|---|---|---|---|---|---|---|---|---|---|
| Value (meV) | 1.29(6) | 1.13(4) | 0.58(6) | 0.45(4) | 0.015(8) | 1.11(5) | $0.5 \times 10^{-3}$ | 0.52(8) | 0.46(7) |
| Co-Co distance (Å) | 4.662 | 5.323 | 4.797 | 5.442 | 5.186 | 5.884 | | | |

**References**

1. J. G. Bednorz, and K. A. Müller, Z. Phys. B **64**, 189 (1986).

2. B. Keimer, S. A. Kivelson, M. R. Norman, S. Uchida and J. Zaanen, Nature **518**, 179 (2015)

3. Yoichi Kamihara, Takumi Watanabe, Masahiro Hirano, and Hideo Hosono, J. Am. Chem. Soc. **130**, 3296 (2008).

4. T. J. Williams, A. A. Aczel, M. D. Lumsden, S. E. Nagler, M. B. Stone, J.-Q. Yan, and D. Mandrus, Phys. Rev. B **92**, 144404 (2015).

5. Qiang Zhang, Feng Ye, Wei Tian, Huibo Cao, Songxue Chi, Biao Hu, Zhenyu Diao, David A. Tennant, Rongying Jin, Jiandi Zhang, and Ward Plummer, Phys. Rev. B, **95**, 220403(R) (2017).

6. Sergey Borisenko, Daniil Evtushinsky, Quinn Gibson, Alexander Yaresko, Timur Kim, MN Ali, Bernd Buechner, Moritz Hoesch, Robert J Cava, arXiv :1507.04847 (2015).

7. Von Hj. Mattausch, and Hk. Muller-Buschbaum, Anorg. Allg. Chem. **386**, 1 (1971).

8. K. Boulahya, M. Parras, A. Vegas, and J. M. González-Calbet, Solid State Sciences **2**, 57 (1998).

9. N. W. Ashcroft and N. D. Mermin, Solid State Physics, page 673, Holt, Rinehart and Winston, New York (1976).

10. K. Boulahya, M. Parras, J. M. González-Calbet, U. Amador, J. L. Martínez, and M. T. Fernández-Díaz, Chem. Mater. **18**, 3898 (2006).

11. R. Jin, Hao Sha, P. G. Khalifah, R. E. Sykora, B. C. Sales, D. Mandrus, and Jiandi Zhang, Phys. Rev. B **73**, 174404 (2006).

12. A. P. Ramirez, MRS Bull. **30**, 447 (2005).

13. A. A. Zvyagin, Low Temperature Phys. **39**, 901 (2013).

14. L. J. de Jongh and A. R. Miedema, Adv. Phys. **23**, 1 (1974).

15. Peter L. Russo, Jun Sugiyama, Jess H. Brewer, Eduardo J. Ansaldo, Scott L. Stubbs, Kim H. Chow, Rongying Jin, Hao Sha, and Jiandi Zhang, Phys. Rev. B **80,** 104421 (1609).

16. H. -J. Koo, K. -S. Lee, and M. -H. Whangbo, Inorg. Chem., **45** (26), 10743 (2006).

17. M. -H. Whangbo, D. Dai, and H, -J. Koo, Solid State Sci. **7**, 827 (2005).

18. M, -H. Whangbo, H, -J. Koo, and D. Dai, J. Solid State Chem. **176**, 417 (2003).

19. H.- J. Koo, M., H. Whangbo, and K. -S. Lee, J. Solid State Chem. **169**, 143 (2002).

20. See Supplemental Material at XXXXX for details of experimental methods.

21. See Supplemental Material at XXXXX for details of crystalline and magnetic structure determination with neutron-diffraction.

22. B. C. Chakoumakos, H. Cao, F. Ye, A. D. Stoica, M. Popovici, M. Sundaram, W. Zhou, J. S. Hicks, G. W. Lynn and R. A. Riedel, J. Appl. Crystallogr. **44**, 655 (2011).

23. P. Hidnert, Journal of Research of the National Bureau of Standards **52**, 311 (1954).

24. A. S. Wills, Physica B **276–278**, 680 (2000).

25. J.M. Perez-Mato, S.V. Gallego, E.S. Tasci, L. Elcoro, G. de la Flor, and M.I. Aroyo, Annu. Rev. Mater. Res. **45**, 217 (2015).

26. Michel Pouchard, Antoine Villesuzanne, and Jean-Pierre Doumerc, Comptes Rendus Chimie **6**, 135 (2003).

27. Matthieu Kauffmann, Olivier Mentré, Alexandre Legris, Sylvie Hébert, Alain Pautrat, and Pascal Roussel, Chem. Mater. **20**, 1741 (2008).

28. See Supplemental Material at XXXXX for details of spin wave dispersion determination.

29. S. Toth and B. Lake, J. Phys.: Condens. Matter **27**, 166002 (2015).

30. M. C. Cross and D. S. Fisher, Phys. Rev. B **19**, 402 (1979).

31. Y. Tokura, S. Seki and N. Nagaosa, Rep. Prog. Phys. **77**, 076501 (2014).

32. P. W. Anderson, Phys. Rev. **79**, 350 (1950).

33. See Supplemental Material (Tables SIV and SV) at XXXXX for details of interatomic distance, angles of the $CoO_4$ tetrahedra and geometrical parameters of the SSE pathways.

34. See Supplemental Material at XXXXX for details of possible magnetic exchange pathways.

35. R. Sinclair, H. D. Zhou, M. Lee, E. S. Choi, G. Li, T. Hong, and S. Calder, Phys. Rev. B **95**, 174410 (2017).

36. Angela Möller, Ngozi E. Amuneke, Phillip Daniel, Bernd Lorenz, Clarina R. de la Cruz, Melissa Gooch, and Paul C. W. Chu, Phys. Rev. B **85**, 214422 (2012).

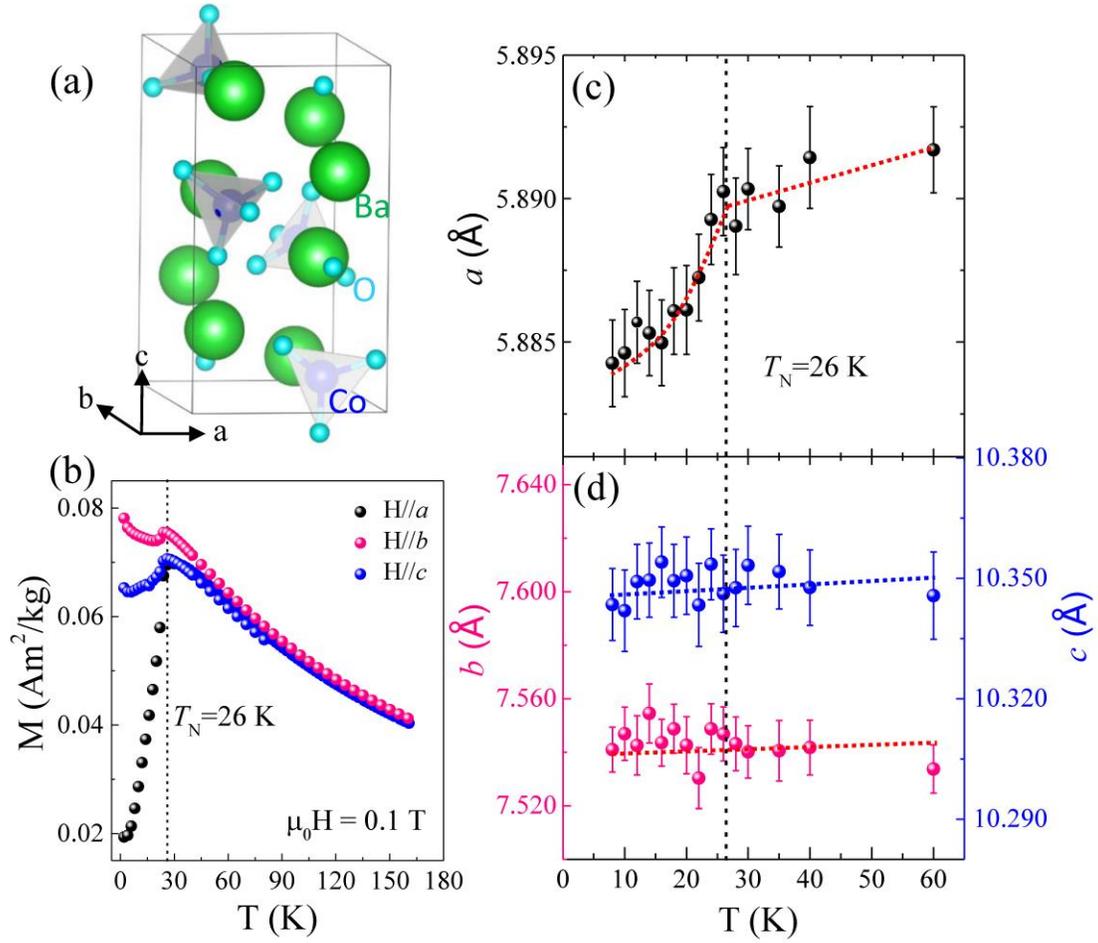

**Fig. 1.** Structural information of $Ba_2CoO_4$. (a) A 3D view of the monoclinic unit cell for $Ba_8Co_4O_{16}$ (simplified as $Ba_2CoO_4$). (b) *T*-dependence of the magnetization curves for $Ba_2CoO_4$ in a field of 0.1 T applied parallel to the crystalline *a*-, *b*- and *c*-axes. (c) and (d) *T*-dependence of the lattice constants *a*, *b* and *c*. The vertical dashed line shows the location of AFM transition temperature $T_N$.

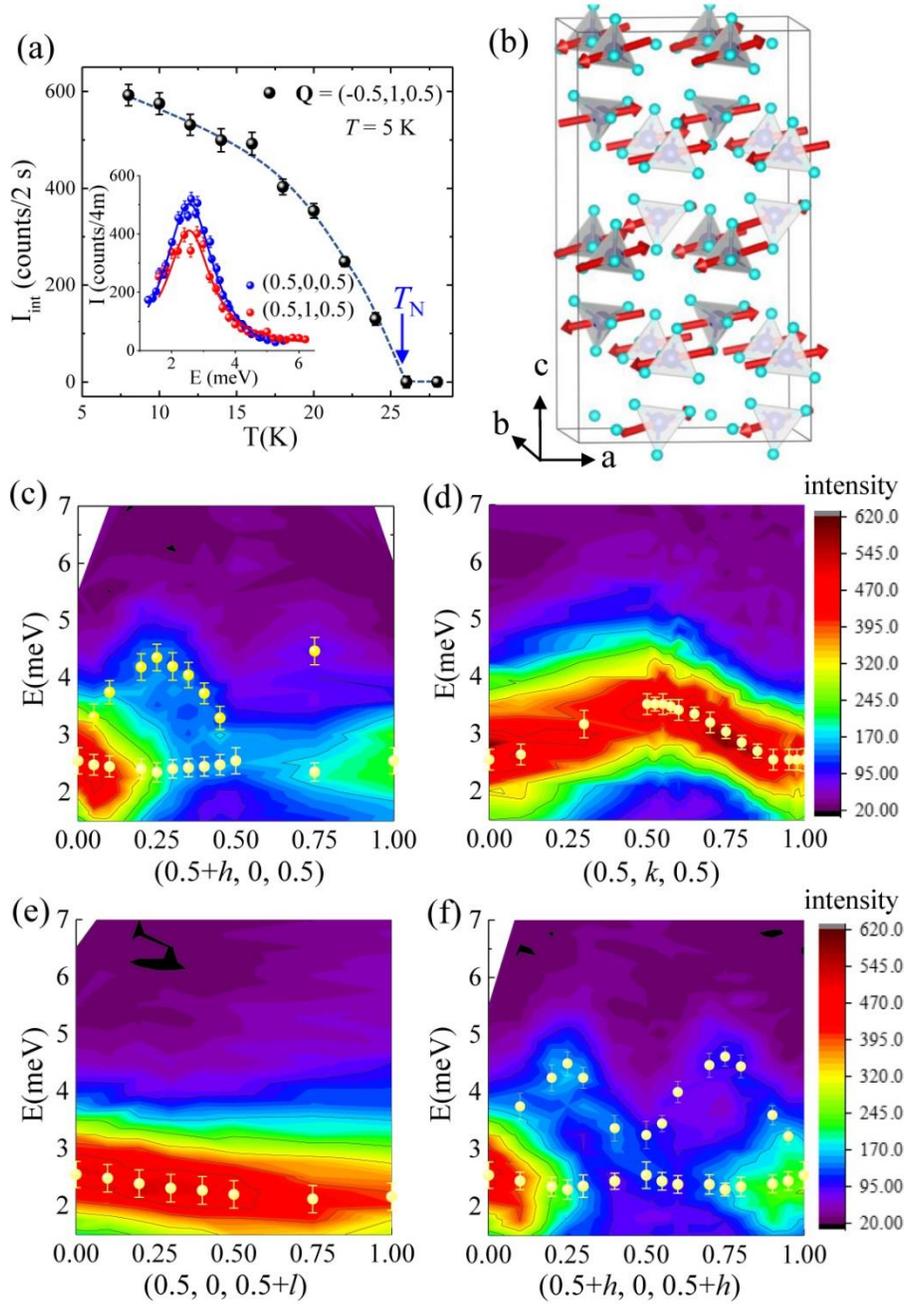

**Fig. 2.** (a) Order parameter of magnetic (-0.5,1,0.5) peak. The inset shows the spin gap at two magnetic zone centers (0.5,0,0.5) and (0.5,1,0.5). (b) The 3D graphic representation of the determined magnetic structure of $Ba_2CoO_4$ at 5 K within $2a \times 1.3b \times 2c$ unit cells (note that one magnetic unit cell is $2a \times b \times 2c$ unit cells). (c)-(f) Experimental $S(Q, E)$ contour plots along the $[H\,0\,0]$, $[0\,K\,0]$, $[0\,0\,L]$ and $[H\,0\,H]$ directions in the r.l.u. The dot symbols show the experimental spin-wave dispersions obtained by the fits to the raw data (see details in the main text).

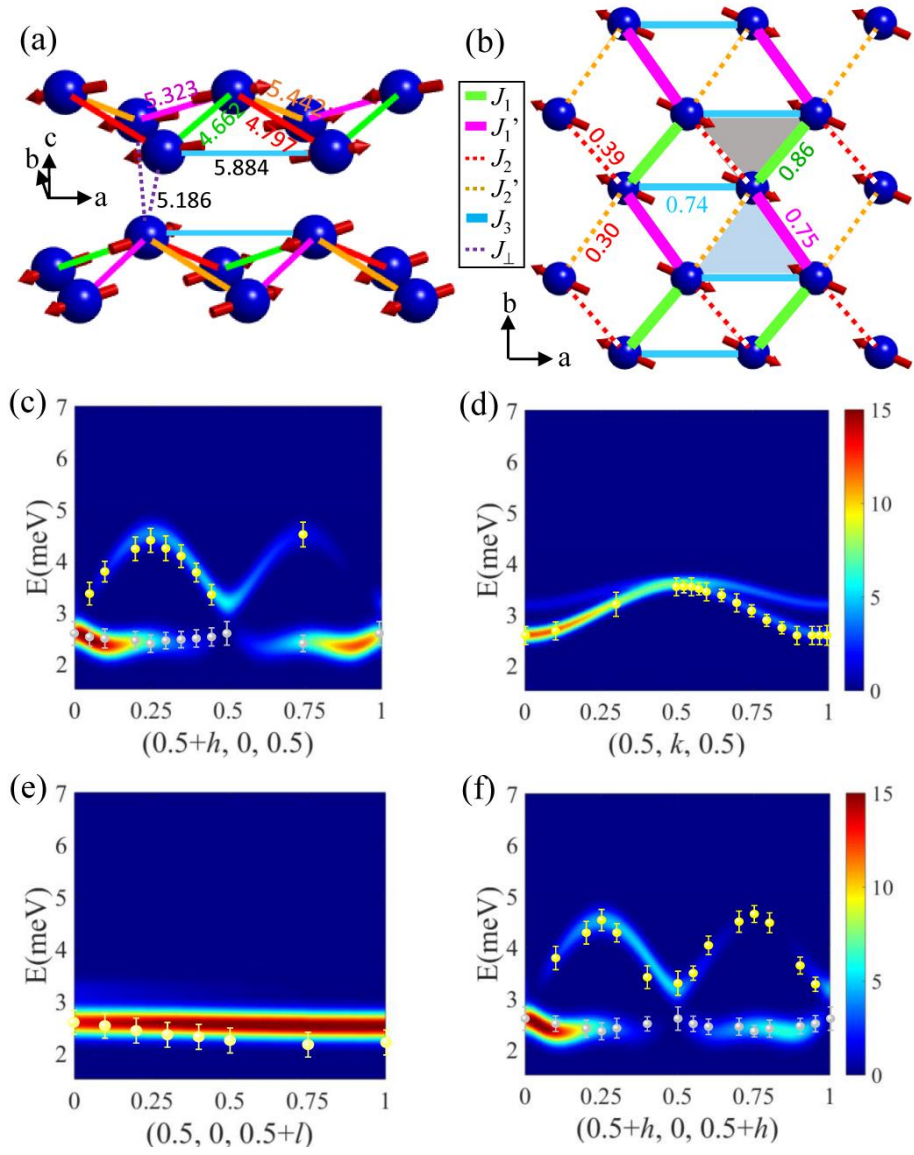

**Fig. 3.** (a) 3D illustration of Co-Co interaction network with spin configuration of two quasi $CoO_4$ PLSs [see Fig. 2(b)] with the Co NN distances. (b) The *ab*-plane projection of one quasi $CoO_4$ PLS showing Co-Co interactions, spin configurations with NN exchange parameters, buckled zigzag chains along the *b* axis, as well as the formed two triangular sublattices. (c)-(f) Simulated $S(Q, E)$ spectra along the [$H$ 0 0], [0 $K$ 0], [0 0 $L$] and [$H$ 0 $H$] directions in the r.l.u. using the magnetic exchange parameters obtained from the fits to the experimental SW dispersion and simulated intensity.

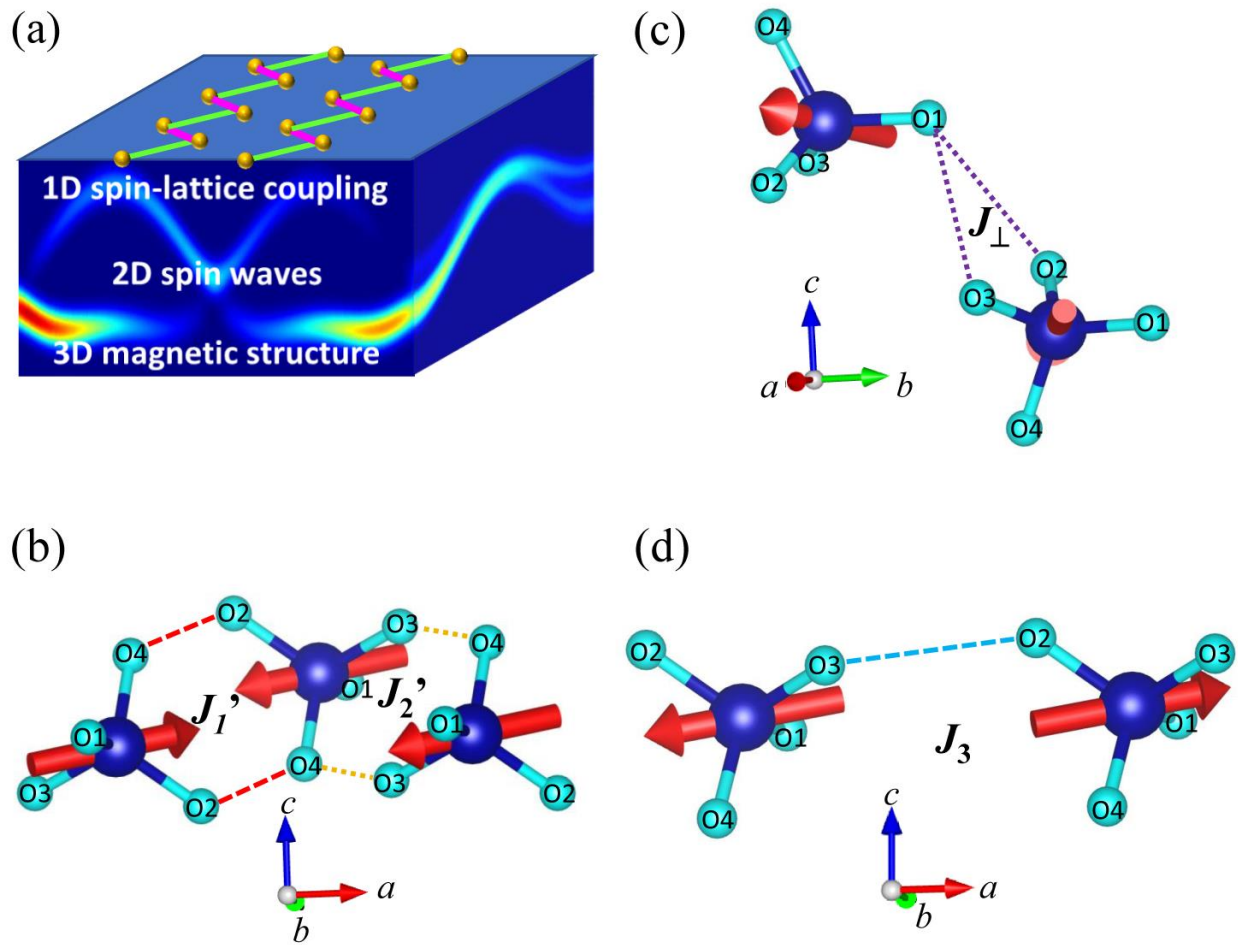

**Fig. 4.** (a) An illustrated view of different dimensional magnetism in $Ba_2CoO_4$. The zigzag magnetic chains are schematically shown. Geometrical representation of indirect spin exchange interaction paths of Co-O---O-Co associated with (b) inter-PLS NN interaction $J_\perp$, (c) intra-PLS NN interaction $J_1'$ and $J_2'$, (c) and (d) intra-PLS NNN interaction $J_3$ in $Ba_2CoO_4$.

Supplemental material for

**Anomalous Magnetic Behavior in Ba$_2$CoO$_4$ with Isolated CoO$_4$ Tetrahedra**

Qiang Zhang[1,2], Guixin Cao[1], Feng Ye[2], Huibo Cao[2], Masaaki Matsuda[2], David A. Tennant[2], Songxue Chi[2], Steven Nagler[2], W.A. Shelton[3], Rongying Jin[1], Ward Plummer[1], Jiandi Zhang[1*]

[1]Department of Physics and Astronomy, Louisiana State University, Baton Rouge, Louisiana 70803, USA

[2]Neutron Scattering Division, Oak Ridge National Laboratory, Oak Ridge, Tennessee 37831, USA

[3]Cain Department of Chemical Engineering, Louisiana State University, Baton Rouge, Louisiana 70803, USA

**Including**

1. **Methods**
2. **Crystalline and magnetic structure of Ba$_2$CoO$_4$**
3. **Spin waves dispersion of Ba$_2$CoO$_4$**
4. **Possible magnetic exchange pathways**

**Note 1: Methods**

**Sample preparation and magnetization measurements**

The $Ba_2CoO_4$ single crystal was grown using the floating zone method with an NEC SC-M15HD image furnace [17]. Magnetization was measured using a superconducting quantum interference device.

**Elastic and inelastic neutron scattering measurements**

Two series of single-crystal elastic neutron scattering experiments to determine the crystal structure and magnetic structure were conducted at the four-circle single crystal diffractometer HB3A with a wavelength of 1.524 Å from the bent Si-331 [22] with and without ~1.4% λ/2 contamination at the High Flux Isotope Reactor (HFIR) Oak Ridge National Laboratory, USA. A single piece of crystal was aligned in the (H,0,L) plane firstly and then (H,0,H)-(0,K,0) plane for spin wave measurements on the HB3 spectrometer at HFIR with a fixed final energy ($E_f$ = 14.7 meV) and HB1 spectrometer at HFIR with a fixed final energy ($E_f$ = 13.5 meV), respectively, to obtain the comprehensive spin-wave dispersion along the three principle axes and the [$H$ 0 $H$] diagonal direction.

**Fits and simulations to spin waves**

The fits to experimental spin waves and the simulation on the convoluted spin waves spectra were performed using classical and quasi classical numerical methods by SpinW package [29]. The constant-Q energy scans were fitted using one (or two) Lorentz functions to obtain both the spin wave dispersion and intensity. The fits to the experimental spin wave dispersion and intensity yield the magnetic exchange constants. To show the simulated spin waves dispersion and intensity well, we used a higher instrumental resolution of 0.4 meV than the experimental 1.2 meV to obtain the convoluted spin waves spectra as shown in Fig. 3 (c-f) in the main text.

**Note 2: Crystalline and magnetic structure of $Ba_2CoO_4$**

**Crystalline structure:** Single crystal neutron diffraction was performed on the same piece of $Ba_2CoO_4$ crystal of at HB-3A at HFIR, ORNL to determine the crystalline and magnetic structure. We employed a wavelength of 1.524 Å involving ~1.4% $\lambda/2$ contamination from the Si-220 monochromator in high resolution mode (bending 150) [22]. After that, using the PG filter available at HB-3A to filter the $\lambda/2$ contamination, we used the same wavelength to remeasure the data sets. The refinement on the two data sets reveal consistent results on the structure and magnetic structure.

To investigate the crystal structure, we collected the full data set of nuclear peaks to a high-Q region at 5 K. The lattice parameters and the observed nuclear reflections reveal a monoclinic structure with the space group $P2_1/n$ (No. 14) at 5 K as illustrated in Fig. 1 (a) in the main text, and we found there is no structural transition from RT down to 5 K. In addition, we observed a twinning along the *a/c* axis in the crystal resulting from a monoclinic lattice distortion in the *ac* plane. A comparison of the observed and calculated values of the squared structure factors from nuclear peaks at 5 K is shown in Fig. S1 where $R_f$ = 4.63% and $\chi^2$ = 0.275. The refined lattice parameters, atomic positions, and reliability factors from our data taken at 5 K are summarized in Table SI.

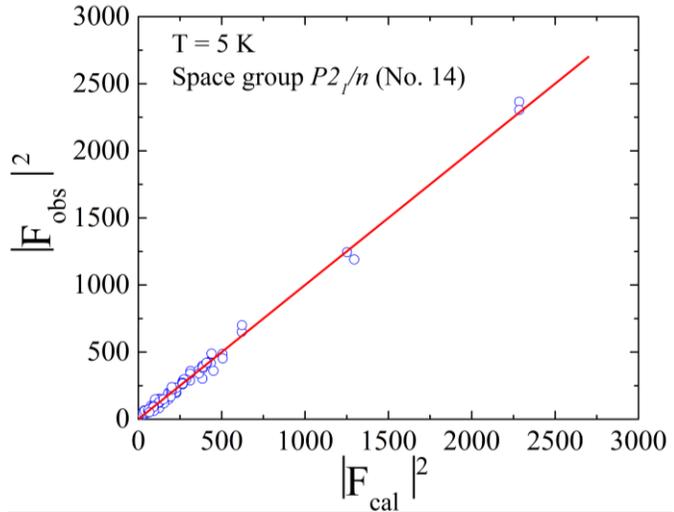

**Fig. S1**. Comparison of the observed and calculated squared structure factors for the nuclear peaks at 5 K using the incident beam with the $\lambda/2$ contamination in $Ba_2CoO_4$.

**Table SI.** Atomic parameters determined from the refinement on the integrated intensities of nuclear peaks on a single crystal of $Ba_2CoO_4$ at 5 K. The space group, $P2_1/n$ (No. 14) has been identified with lattice parameters, $a$ = 5.884 (5) Å, $b$ = 7.540 (3) Å, $c$ = 10.343 (6) Å, $\alpha = \gamma = 90°$, $\beta$ = 90.52 (7)°

|  | Wyckoff site | $x$ | $y$ | $z$ | B-factors | RF-factor | $\chi^2$ |
|---|---|---|---|---|---|---|---|
| Ba1 | 4e | 0.761 (5) | 0.852 (7) | 0.084(5) | 1.872(8) | | |
| Ba2 | 4e | 0.246(7) | 0.492(4) | 0.194 (3) | 1.872(8) | | |
| Co1 | 4e | 0.742(4) | 0.279(3) | 0.077(8) | 1.029(6) | | |
| O1 | 4e | 0.782(6) | 0.507(3) | 0.081(2) | 1.530(4) | | |
| O2 | 4e | 0.520(4) | 0.207(6) | 0.175(5) | 1.530 (4) | | |
| O3 | 4e | 0.990(5) | 0.172(5) | 0.136(3) | 1.530 (4) | | |
| O4 | 4e | 0.708(6) | 0.185(4) | 0.925(3) | 1.530(4) | | |
| Reliable factors | | | | | | 3.76% | 0.275 |

**Magnetic structure:** Below $T_N \approx 26$ K, magnetic Bragg peaks appear with the propagation vector $k$ = (-0.5,0,0.5). Fig. S2(a) displays rocking curves of the representative (-0.5,1,0.5) magnetic peak at 8, 22 and 26 K. The order parameter of (-0.5,1,0.5) indicates that it disappeared at $T_N$ as shown in Fig. 2(a) in the main text. To investigate the magnetic structure, a full data set with propagation vector $k$ = (-0.5,0,0.5) was collected at 5 K. The SARAH representational analysis program [24] was employed to determine the symmetry of the allowed magnetic structures. We summarize the basis vectors of the allowed magnetic structures in Table SII. There are four coordinates or spin sites, i.e., Co1, Co2, Co3 and Co4, of crystallographically equivalent Co sites. Caution has been taken in determining the magnetic structure. We used two methods to refine the magnetic structures carefully on the two data sets with and without λ/2 contamination, and the twin effect was also considered. In the first method, we obtained the structural factors from refining the integrated intensities of the nuclear peaks firstly and then used them to refine the integrated intensities of the magnetic peaks solely by creating PCR file *via* the SARAH program. The other method used one magnetic unit cell ($2a \times b \times 2c$) to re-index the nuclear and magnetic peaks and refine both nuclear and magnetic peaks to get the structural factors and magnetic structure simultaneously. In this way, the PCR file for refinement was created by the Bilbao Crystallographic Server [25]. Both ways yielded the same magnetic structure. The integrated intensities of all the nuclear and magnetic reflections can be best fitted employing the $\Gamma 3$ magnetic structure with the fits yielding $|m_a| = 2.377(6)$ $\mu_B$, $|m_b| = 1.128(5)$ $\mu_B$, and $|m_c| = 0.586(7)$ $\mu_B$. Fig. S2 (b) shows the comparison between observed and calculated values of the squared structure factor for re-indexed nuclear and magnetic peaks using one magnetic unit cell where we find a $R_f$ of 4.43% and $\chi^2$ of 0.4. The magnetic moment of $Co^{4+}$, mainly confined in the *ab* plane, points to the *a* axis, with a canting to the *b* axis $\approx 25°$ and also a slight canting to the *c* axis $\approx 13°$, as illustrated in Fig. 2 (b).

**Table SII.** The symmetry-allowed basis vector [$m_x$, $m_y$, $m_z$] for the space group $P2_1/n$ (No. 14) with $k = (-0.5, 0, 0.5)$ at 5 K in $Ba_2CoO_4$. Co1: (x, y, z), Co2: (-x+1/2, y+1/2, -z+1/2), Co3: (-x, -y, -z), Co4: (x+1/2, -y+1/2, z+1/2).

| IR | $\Gamma 1$ | $\Gamma 2$ | $\Gamma 3$ | $\Gamma 4$ |
|---|---|---|---|---|
| Co1 | [mx my mz] | [mx my mz] | [mx my mz] | [mx my mz] |
| Co2 | [-mx my -mz] | [-mx my -mz] | [mx -my mz] | [mx -my mz] |
| Co3 | [mx my mz] | [-mx -my -mz] | [mx my mz] | [-mx -my –mz] |
| Co4 | [-mx my -mz] | [mx -my mz] | [mx -my mz] | [-mx my -mz] |

Note that our magnetic structure determined from the refinement on many magnetic Bragg peaks of single crystals is different from the previous report obtained by the refinement on very few magnetic peaks with powder neutron diffraction [10]. The moment mainly along *c* axis reported previously [10] can be excluded easily and safely. It is also worthwhile pointing out that in addition to a canting toward the *b* axis, there is a weak canting along the *c* axis since the restriction of moment in the *ab* plane lead to a worse refinement and failed to refine a few strong

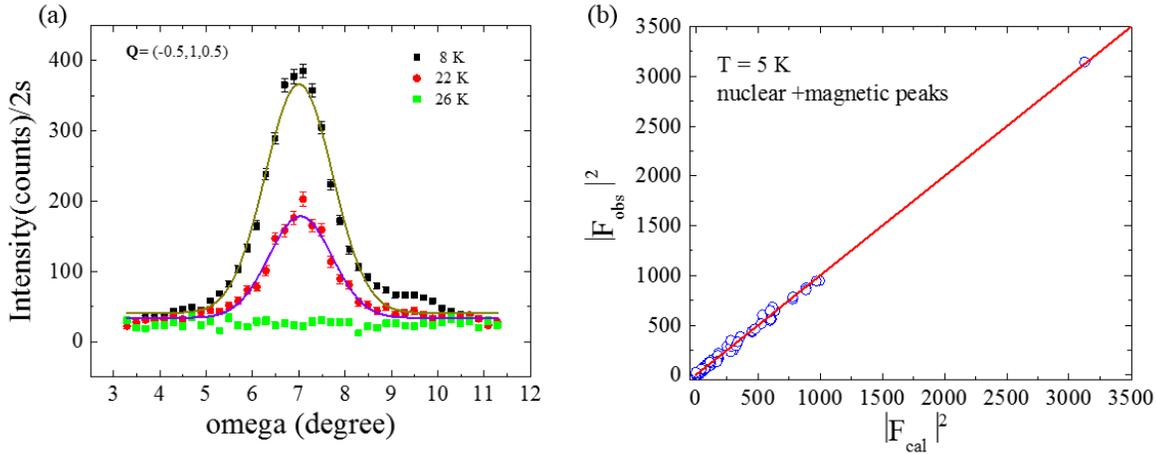

**Fig. S2.** (a). Rocking-curve scans of (-0.5,1,0.5) magnetic peak at 8, 22 and 26 K. (b). Comparison of the observed and calculated squared structure factors of the nuclear and magnetic peaks using Γ3 magnetic configuration at 5 K in $Ba_2CoO_4$.

magnetic peaks, such as (1.5,0,-1.5).

**Note 3: Spin waves dispersion of $Ba_2CoO_4$**

To map out the spin waves dispersion, constant-Q energy scans at various Q's are conducted along the high-symmetric *H, K, L* and the [*H* 0 *H*] directions. Representative scans along the [*H H* 0] and *L* directions are shown in Fig. S3. There are two branches of spin waves along the [*H H* 0] direction and merged as one broad branch of spin waves along the *L* direction. As *H* moves from the magnetic zone center (1,0,1) to the boundary (1.25,0,1.25), both the spin waves intensities and peak positions vary for both branches. In contrast, as *L* increases, the spin waves intensity remains unchanged and there is negligible change in peak positions, indicative of an almost flat spin wave along the *L* direction. The solid lines are the best fits using the Lorentz functions to get the peak positions while the obtained experimental spin-waves dispersions are shown as the open symbols in Fig. 3(c)-(f).

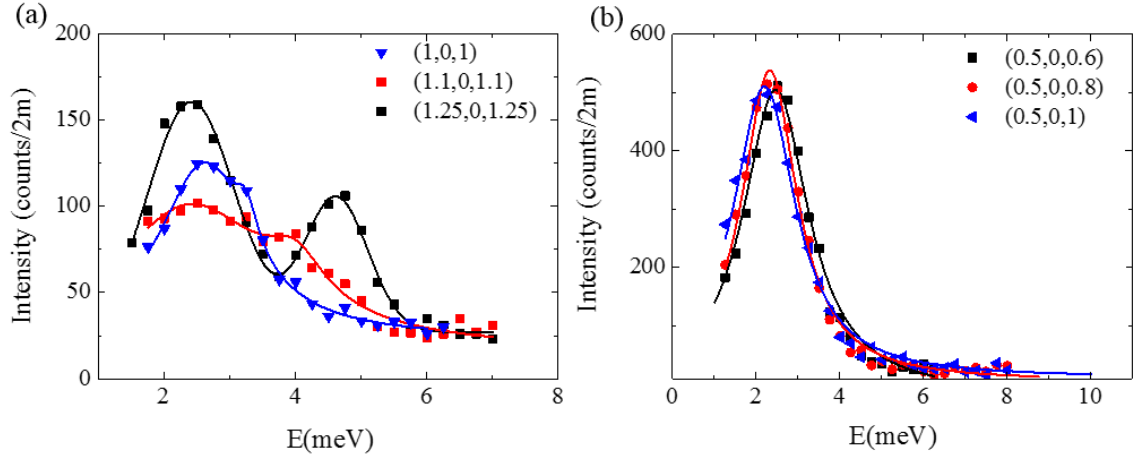

**Fig. S3.** Representative spin waves measurements by conducting constant-Q energy scans at different Q's along (a) [*H* 0 *H*] and (b) *L* directions in reciprocal lattice unit (*r.l.u.*) at 5 K for $Ba_2CoO_4$. Solid lines in (a) and (b) are the fits using two and one Lorentz functions, respectively.

**Note 4: Possible magnetic exchange pathways**

As discussed in the main text, the indirect spin exchange interactions may take place through the paths Co-O···O-Co or even Co-O-Ba-O-Co. The super-superexchange interaction (SSE) by the paths Co-O···O-Co was proposed previously [16] as reported in other compounds with isolated octahedra in $NaFeP_2O_7$ [17] or isolated tetrahedra in $Ba_3Cr_2O_8$, $Ni_4FeO_4$ [16-18], and sulfides $BaLn_2MnS_5$ (Ln=La,Ce, Pr) [19]. In these examples, the sign and the magnitude of such SSEs *via* M-L···L-M (M=transition metal element; L=O, S, Cl, or Br) are found to be governed not by the direct M-M distance, but by the overlap of the magnetic orbitals of the $ML_4$ unit, especially the overlap of their *p*-orbital tails of the non-bonding L···L contacts in the vicinity of the van der Waals distance. The L···L distances, the M–L···L or L···L-M angles also play a critical role in determining the overlap of the L *p*-orbital tails. The SSE interactions usually become stronger when the spin units are planar and rectangular since the shape and orientation of the L $p$-orbital show tendency to overlap due to short L⋯L distances and the large M–L⋯L angles.

In $Ba_2CoO_4$, as shown in Fig. 4 (b-d) and summarized in Table SIII from our Rietveld refinements of neutron diffraction, there are two reasonable Co-O⋯O-Co paths associated with $J_\perp$, $J_1'$ and $J_2'$ to create the SSEs. For one pathway Co-$O_1$⋯$O_3$-Co of $J_\perp$, the $O_3$-$O_1$ distance in is longer than those on $J_1'$ or $J_2'$ and the angle ∠Co-$O_3$⋯$O_1$ of $J_\perp$ is close to 90º, both of which yield a very weak magnetic interaction $J_\perp$ via pathway Co-$O_1$⋯$O_3$-Co. As for another pathway Co-$O_1$⋯$O_2$-Co of $J_\perp$, although the $O_2$⋯$O_1$ distance is slightly shorter than those on $J_1'$ or $J_2'$, both ∠Co-$O_2$⋯$O_1$ and ∠$O_2$⋯$O_1$-Co are smaller than those on $J_1'$ or $J_2'$, likely yielding a weaker spin exchange interaction via the Co-$O_2$⋯$O_1$-Co path as well. Thus, the overall interchain $J_\perp$ via these two pathways can be much weaker than $J_1'$ or $J_2'$. When one compares the paths for $J_3$ and $J_\perp$, the $O_3$-$O_2$ distance for $J_3$ is much shorter and the ∠Co-$O_3$⋯$O_2$ and ∠$O_3$⋯$O_2$-Co angles are much larger than corresponding ones for $J_\perp$, which could be a cause why $J_3$ is much larger than $J_\perp$. Therefore, it is understood that the overall interchain $J_\perp$ is much weaker than $J_1'$, $J_2'$ or $J_3$ within the frame work of Co-O⋯O-Co exchange pathway although the direct Co-Co distance on $J_\perp$ is shorter.

**Table SIII.** Interatomic distances and angles of the $CoO_4$ tetrahedra and geometrical parameters of the SSE paths Co-O···O-Co associated with $J$ in $Ba_2CoO_4$.

| Interatomic distances and angles of the $CoO_4$ tetrahedra | | | |
|---|---|---|---|
| Bond | Distance (Å) | Bond | angle (°) |
| Co-O1 | 1.739 | O1-Co-O2 | 113.667 |
| Co-O2 | 1.717 | O1-Co-O3 | 109.71 |
| Co-O3 | 1.772 | O1-Co-O4 | 115.77 |
| Co-O4 | 1.74 | O2-Co-O3 | 106.98 |
| | | O2-Co-O4 | 107.4 |
| | | O3-Co-O4 | 102.4 |
| Label | Geometrical parameters of the SSE paths Co-O···O-Co associated with $J$ | | | | | |
| | d(O3-O1) | Co-O3···O1 | O3···O1-Co | d(O2-O1) | Co-O2···O1 | O2···O1-Co |
| $J_\perp$ | 3.562 | 93.7 | 107.4 | 3.186 | 109.1 | 122.1 |
| | d(O4-O2) | Co-O4···O2 | O4···O2-Co | d(O2-O4) | Co-O2···O4 | O2···O4-Co |
| $J_1'$ | 3.399 | 131.147 | 114.3 | 3.399 | 114.3 | 131.147 |
| | d(O3-O4) | Co-O3···O4 | O3···O4-Co | d(O4-O3) | Co-O4···O3 | O4···O3-Co |
| $J_2'$ | 3.285 | 138.42 | 117.0 | 3.285 | 117.0 | 138.42 |
| | d(O3-O2) | Co-O3···O2 | O3···O2-Co | | | |
| $J_3$ | 3.146 | 144.19 | 136.363 | | | |

We now discuss another likely SSE pathway, Co-O-Ba-O-Co, in $Ba_2CoO_4$. We summarize the bond length and angles in Table SIV based on our Rietveld refinements of the neutron diffraction patterns and present the Co-O-Ba-O-Co pathways associated with all the exchange constants in Fig. S4(a-d). It turns out that the magnitude of exchange constants is not proportional to the total path of the Co-O-Ba-O-Co, as is expected for the indirect magnetic interaction. Thus, the total number of the pathways and bond angles need to be considered. For the exchange pathway Co-O-Ba-O-Co, the filled outermost 5 $p$ orbital of $Ba^{2+}$ combines the 2$p$ orbitals of the left $O^{2-}$ ion to form one molecular orbital so that the spin on left $Co^{4+}$ can be transferred to the molecular orbital. Correspondingly, another magnetic molecular orbital is also formed by the filled outermost $5p$ orbital $Ba^{2+}$ and the $2p$ orbitals of the right $O^{2-}$ ion to transfer the spin on the right $Co^{4+}$. Therefore, the spin exchange may be built due to the overlapped two molecular orbitals by the bridging $Ba^{2+}$ via Co-O-Ba-O-Co, as seen in $R$Cr(BO$_3$)$_2$ ($R$ = Y and Ho) [35] and $A$Ag$_2M$[VO$_4$]$_2$ ($A$ = Ba, Sr; $M$ = Co, Ni) [36] involving this type of pathway through nonmagnetic cations. Compared to other exchange constants, the magnitude of $J_\perp$ is much weaker, and the following mechanisms probably need to be considered. Firstly, there is only one Co-O-Ba-O-Co path on $J_\perp$. In contrast, there are two paths on $J_1'$, $J_2'$ that could result in larger overall magnetic interactions. Additionally, the ∠O$_2$-Ba$_2$-O$_3$ angle of $J_\perp$ is 96.8°, close to 90°, which may tend to result in a weaker magnetic interaction between two magnetic molecular orbitals. This is supported by the much larger $J_3$ with only one Co-O-Ba-O-Co path but a much larger ∠O$_4$-Ba$_1$-O$_3$ angle (120.2°). However, such SSE interaction through the nonmagnetic cation is supposed to be very weak and usually ferromagnetic type [35,36], in sharp contrast to the strong AFM interactions in Ba$_2$CoO$_4$. Thus, we conclude that the dominant pathway in Ba$_2$CoO$_4$ may be Co-O⋯O-Co rather than Co-O-Ba-O-Co.

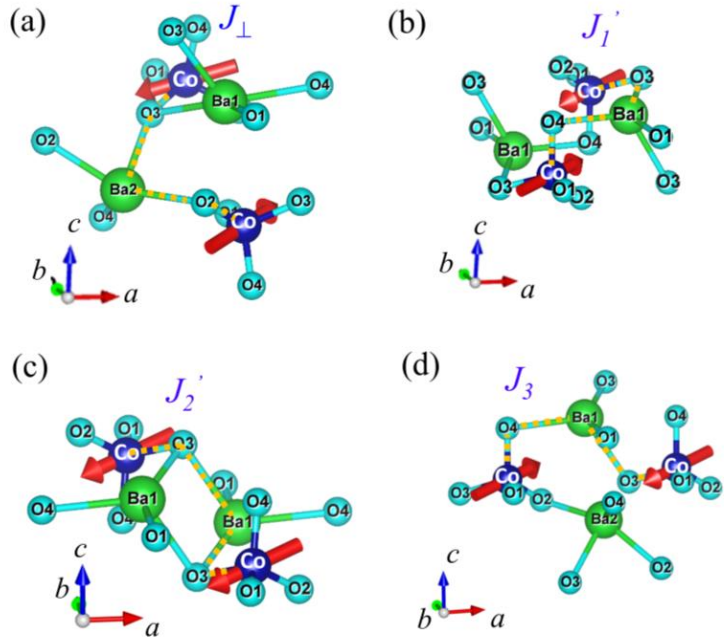

**Fig. S4.** Geometrical representation of exchange paths of Co-O-Ba-O-Co associated with (a) $J_\perp$, (b) $J_1'$, (c) $J_2'$, and (d) $J_3$ in $Ba_2CoO_4$. Different colors of the dotted lines indicate different exchange pathways. However, the equivalent pathways are omitted.

**Table SIV.** Geometrical parameters of the exchange paths Co-O-Ba-O-Co associated with $J$ in $Ba_2CoO_4$.

| Label | Geometrical parameters of the exchange paths Co-O-Ba-O-Co associated with $J$ | | | |
|---|---|---|---|---|
| $J_\perp$ | d(Co-O2-Ba2-O3-Co) | ∠Co-O2-Ba2 | ∠O2-Ba2-O3 | ∠Ba2-O3-Co |
| (1 path) | 8.86 | 105.02 | 96.8 | 157.1 |
| $J_1'$ | d(Co-O3-Ba1-O4-Co) | ∠Co-O3-Ba1 | ∠O3-Ba1-O4 | ∠Ba1-O4-Co |
| (2 paths) | 9.1023 | 86.0 | 124.7 | 100.28 |
| $J_2'$ | d(Co-O3-Ba1-O3-Co) | ∠Co-O3-Ba1 | ∠O3-Ba1-O3 | ∠Ba1-O3-Co |
| (2 paths) | 9.081 | 100.67 | 87.79 | 86 |
| $J_3$ | d(Co-O4-Ba1-O3-Co) | ∠Co-O4-Ba1 | ∠O4-Ba1-O3 | ∠Ba1-O3-Co |
| (1 path) | 9.024 | 100.28 | 120.2 | 100.69 |